\documentclass[journal,12pt,draftclsnofoot,onecolumn]{IEEEtran}
\usepackage{mathrsfs}
\usepackage[cmex10]{amsmath}
\usepackage{amsfonts,amssymb}
\usepackage{mdwmath}
\usepackage{amsthm}
\usepackage{subfigure}
\usepackage{epsf}
\usepackage{graphicx}
\usepackage{subfigure}
\usepackage[sort,compress]{cite}
\usepackage{algorithmic}
\usepackage[ruled,vlined]{algorithm2e}
\usepackage{booktabs} 
\usepackage{multirow}
\usepackage{color}            
\usepackage{threeparttable}   
\usepackage{lipsum,multicol}  

\theoremstyle{definition}

\begin{document}
\begin{titlepage}
\begin{center}
\vspace*{-2\baselineskip}
\begin{minipage}[l]{7cm}
\flushleft
\includegraphics[width=2 in]{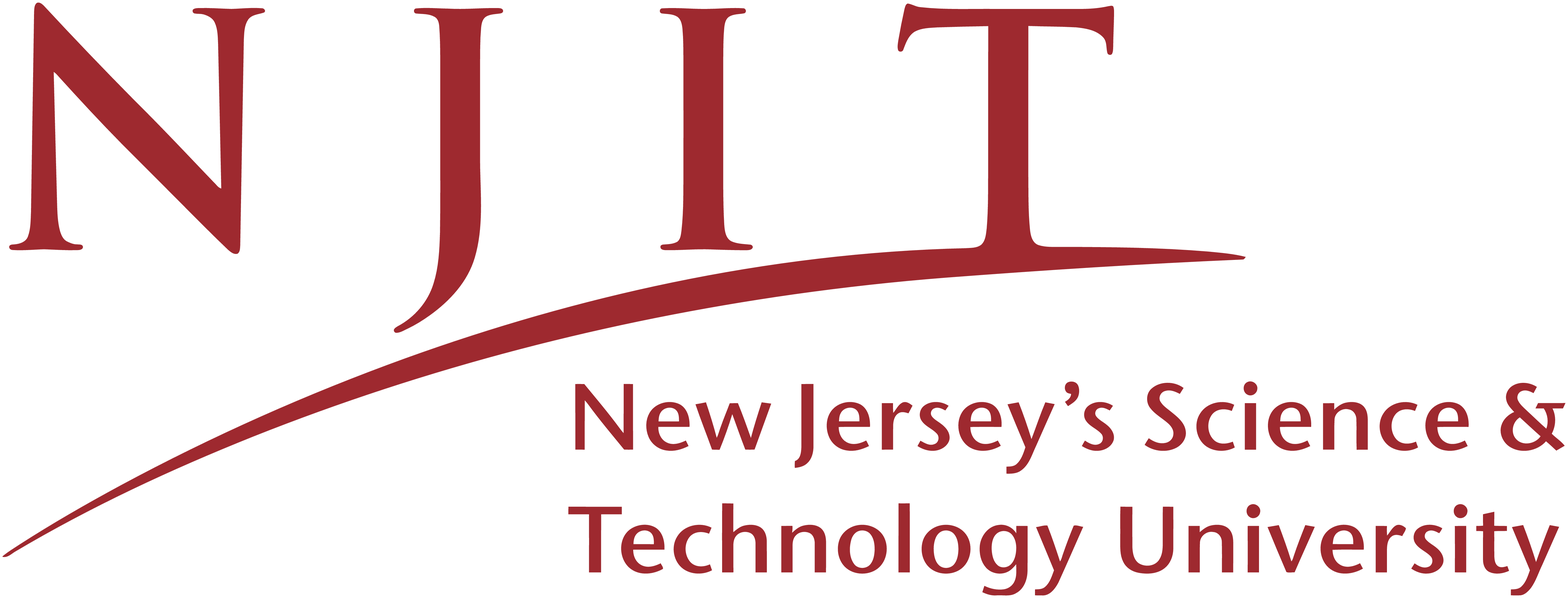}
\end{minipage}
\hfill
\begin{minipage}[r]{7cm}
\flushright
\includegraphics[width=1 in]{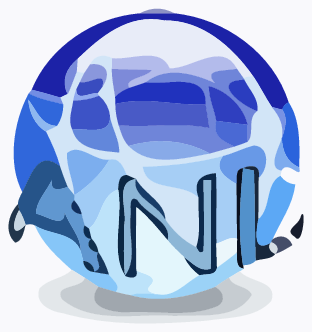}
\end{minipage}
\vfill
\textsc{\LARGE Energy-Aware Virtual Machine Management in Inter-datacenter Networks over Elastic\\[12pt]
Optical Infrastructure}\\
\vfill
\textsc{\LARGE LIANG ZHANG\\[12pt]
\LARGE TAO HAN \\[12pt]
\LARGE NIRWAN ANSARI}\\
\vfill
\textsc{\LARGE TR-ANL-2017-006\\[12pt]
\LARGE October 9, 2017}\\[1.5cm]
\vfill
{ADVANCED NETWORKING LABORATORY\\
 DEPARTMENT OF ELECTRICAL AND COMPUTER ENGINEERING\\
 NEW JERSY INSTITUTE OF TECHNOLOGY}
\vfill
\end{center}

\begin{minipage}[c]{16cm}
\flushleft
\small
{{Citation:}\\
L. Zhang, T. Han, and N. Ansari, ``Energy-Aware Virtual Machine Management in Inter-datacenter Networks over Elastic Optical Infrastructure," \emph{IEEE Trans. on Green Communications and Networking}, DOI: 10.1109/ TGCN.2017.2771724.\\}
{URL:\\
http://ieeexplore.ieee.org/document/8100944/}
\end{minipage}

\end{titlepage}
\title{Energy-Aware Virtual Machine Management \\in Inter-datacenter Networks over Elastic \\Optical Infrastructure}
\author{\IEEEauthorblockN{Liang~Zhang, ~\IEEEmembership{Student member,~IEEE},
                          Tao~Han, ~\IEEEmembership{Member,~IEEE},
                          and~Nirwan Ansari, ~\IEEEmembership{Fellow,~IEEE}}
\thanks{The preliminary idea of this work was presented at IEEE Cloudcom 2015 in Vancouver, Canada,  pp. 440 - 443, Nov. 30 - Dec. 2, 2015. This work was partially supported by National Science Foundation under Grant No. $CNS\!-\!1320468$  and $CNS\!-\!1731675$.}
\thanks{Liang Zhang and Nirwan Ansari are with the Advanced Networking Lab. Dept. Elec. $\&$ Comp. Engrg., New Jersey Institute of Technology, Newark, NJ, 07102, USA. Email: \{lz284, nirwan.ansari\}@njit.edu}
\thanks{Tao Han is with Dept. Elec. $\&$ Comp. Engrg., University of North Carolina at Charlotte, Charlotte, NC, 28223, USA. Email: Tao.Han@uncc.edu}
\thanks{Manuscript received March 13, 2017; revised October 9, 2017; accepted November 2, 2017.}}

%
\maketitle

\begin{abstract}
Datacenters (\emph{DC}s), deployed in a large scale to support the ever increasing demand for data processing applications, consume tremendous energy. Powering DCs with renewable energy can effectively reduce the brown energy consumption. Owing to geographically distributed deployment of DCs, the renewable energy generation and the data processing demands usually vary in different DCs. Migrating virtual machines (\emph{VM}s) among DCs according to the availability of renewable energy helps match the energy demands and the renewable energy generation in DCs, and thus maximizes the utilization of renewable energy. We first elicit the renewable energy-aware inter-datacenter (\emph{inter-DC}) VM migration problem in an inter-DC network over the elastic optical infrastructure, present it as a many-manycast communications problem, and then formulate it as an integer linear programming problem. The objective is to minimize the total cost of the brown energy consumption of DCs in such inter-DC network via VM migration. We use CVX and Gurobi to solve this problem for small network configurations, and we propose a few heuristic algorithms that approximate the optimal solution for large network configurations. Through extensive simulations, we show that the proposed algorithms, by migrating VM among DCs, can reduce up to $19.7\%$ cost of the brown energy consumption.
\end{abstract}

\vspace{-2mm}
\IEEEpeerreviewmaketitle

\section{Introduction}
\IEEEPARstart{C}{loud} infrastructures are widely deployed to support various emerging applications such as: Google App Engine, Microsoft Window Live Service, IBM Blue Cloud~\cite{Sadiku2014}, augmented reality (\emph{AR}), collaborative learning, multimedia recognition and retrieval~\cite{cloud_computing_application2016}. Data center (\emph{DC}) operators such as Amazon and Microsoft provision cloud computing and storage services via Amazon AWS and Microsoft Azure, respectively~\cite{Edge_cloud_computing2017}. Large-scale DCs, which are the fundamental engines for data processing, are the essential elements in cloud computing \cite{Zhangyan_13DC, zhangyan_dc_survey}. Information and Communications Technology (\emph{ICT}) is estimated to be responsible for about $14\%$ of the worldwide energy consumption by 2020 \cite{Pickavet2008}. The energy consumption of DCs accounts for nearly $26\%$ of the total ICT energy consumption \cite{Pickavet2008}. Hence, the energy consumption of DCs becomes an imperative issue.

\begin{figure}[!htb]
    \centering
    \includegraphics[width=1.0\columnwidth]{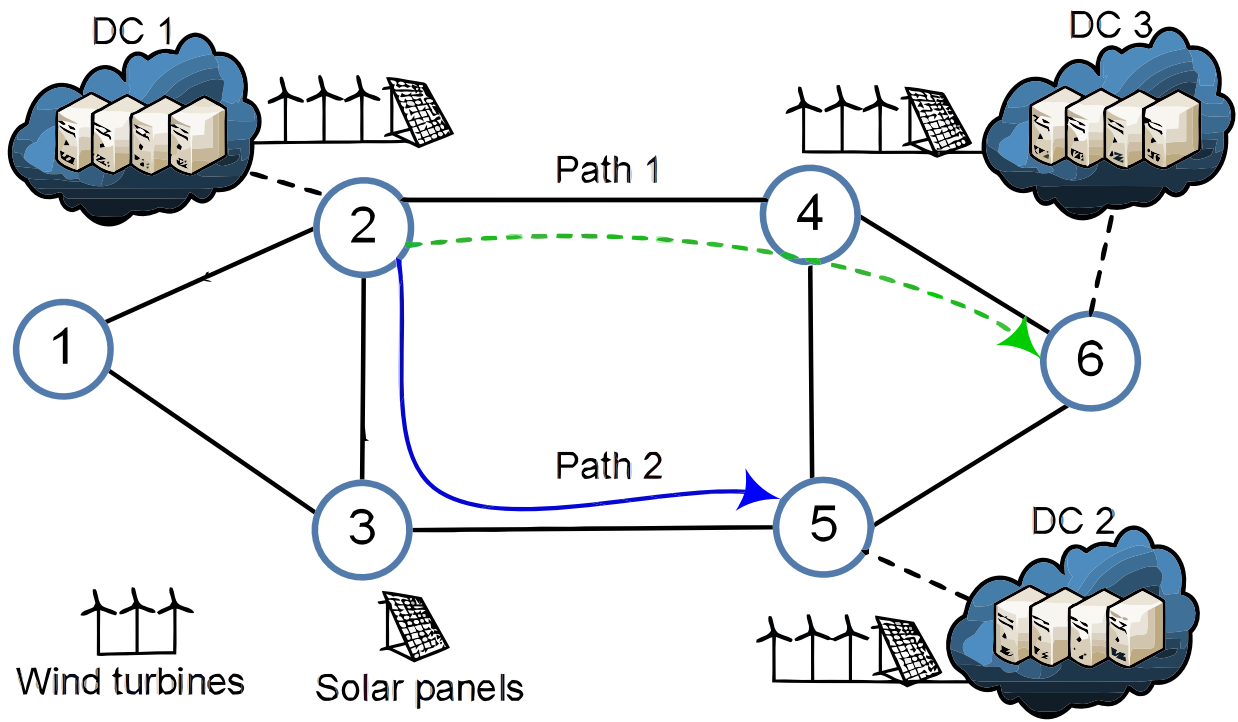}
    \caption{\small Inter-DC architecture.}
    \label{fig:six-cloud}
\end{figure}

Renewable energy, which includes solar and wind, produced $12.7\%$ domestic electricity of the United States in 2011 \cite{Green-cloud2013}, and will be widely adopted to reduce the brown energy consumption of ICT \cite{tao2014_magazine}. Here, brown energy refers to the energy derived from non-renewable resources (e.g., fossil fuel), which generate carbon emissions; green energy refers to the energy derived from renewable resources (e.g., solar, wind, tide, etc.~\cite{Renewable_energy_2017}), which do not generate carbon emissions. For example, Parasol is a solar-powered DC \cite{Goiri_GreenDc}. In Parasol, GreenSwitch, a management system, is designed to manage the work loads and power supplies \cite{Goiri_GreenDc}. The availability of renewable energy varies in different areas and changes over time. The work loads of DCs also vary in different areas and at different time. As a result, the renewable energy availability and energy demands in DCs usually mismatch with each other. This mismatch leads to inefficient renewable energy usage in DCs. To solve this problem, it is desirable to balance the work loads among DCs according to their green energy availability. Although the current cloud computing solutions such as cloud bursting \cite{Wood_cloudNet_2014}, VMware and F5 \cite{VMware_F5} support inter-datacenter (\emph{inter-DC}) virtual machine (\emph{VM}) migration, it is not clear how to migrate VM among renewable energy powered DCs to minimize their brown energy consumption.

Elastic Optical Networks (\emph{EONs}), by employing orthogonal frequency division multiplexing (\emph{OFDM}) techniques, not only provide a high network capacity but also enhance the spectrum efficiency because of the low spectrum granularity \cite{Shieh2007}. The granularity in EONs can be $12.5$ GHz or even smaller \cite{Armstrong2009}. Therefore, EONs are one of the promising networking technologies for inter-DC networks \cite{Develder2012}.

Powering DCs with renewable energy can effectively reduce the brown energy consumption, and thus alleviate green house gas emissions. DCs are usually co-located with the renewable energy generation facilities such as solar and wind farms~\cite{Figuerola_2009}. Transferring renewable energy via the power grid may introduce a significant power loss of up to $15\%$, and it is desirable to use the renewable energy locally (close to the source of the renewable energy generation) rather than transferring the energy back to the power grid~\cite{Green-cloud2013}.

In this paper, we investigate the \emph{r}enewable \emph{e}nergy-\emph{a}ware \emph{i}nter-DC VM \emph{m}igration (\emph{RE-AIM}) problem that maximizes the renewable energy utilization by migrating VMs among DCs.
In this paper, $\upsilon_{max}$ is defined as the ratio of the maximum available network resources that can be used for VM migration over the whole network resources, and it is used to control the available network resources for VM migration. Meanwhile, $(1-\upsilon_{max})$ is the ratio of the network resources, which are occupied by the other traffic such as the background traffic, over the total network resources. Fig. \ref{fig:six-cloud} shows the architecture of an inter-DC network. The vertices in the graph stand for the optical switches in EONs. DCs are connected to the optical switches via IP routers\footnote{In this paper, we focus on the EONs. The design and optimization of the IP networks are beyond the scope of this paper.}. These DCs are powered by hybrid energy including brown energy, solar energy, and wind energy. For example, assume that DC $1$ lacks renewable energy while DC $2$ and DC $3$ have superfluous renewable energy. Some VMs can be migrated out of DC $1$ in order to save brown energy. Because of the background traffic and the limited network resource, migrating VMs using different paths (Path $1$ or Path $2$) has different impacts on the network in terms of the probability of congesting the network. It is desirable to select a migration path with minimal impact on the network.


There are different types of communications. Manycast is the communications from one source node to a set of destination nodes, and this destination node set is a subset of the candidate destination node set \cite{manycast_Salehi_JLT_2015}. In our previous work (\!\cite{Liang_Cloudcom_2015}(Cloudcom paper), \cite{Cloudcom2015_arXiv2}(technical report)), we presented the RE-AIM problem, which has a set of source nodes/DCs and a set of destination nodes/DCs. A DC with sufficient computing resources and abundant renewable energy can accommodate VM migration from multiple DCs, which lack renewable energy; a DC, which lacks renewable energy, can migrate its VMs to multiple DCs to alleviate the mismatch between the energy workload demands and the renewable energy generation. In this work, we need to migrate VMs from a few source DCs to a few destination DCs. Moreover, the migration may happen from one source DC to many DCs or from many DCs to one DC. Then, we have a source node set and a destination node set, thus resulting in many-manycast communications.

To the best of our knowledge, this is the first time many-manycast communication is proposed. Moreover, this is the first work to minimize the cost of the brown energy consumption of DCs in an inter-DC network overlaid on elastic optical infrastructures via VM migration. The rest of the paper is organized as follows. Section \ref{sec:related_work} describes the related work. Section \ref{sec:problem} formulates the RE-AIM problem based on integer linear programming (\emph{ILP}). Section \ref{sec:analysis-algorithms} briefly analyzes the property of the RE-AIM problem, proposes a few heuristic algorithms to solve the problem, and discusses how to implement these algorithms in the inter-DC network. Section \ref{sec:evaluations} demonstrates the viability of the proposed algorithms via extensive simulation results, and compares the performance of the proposed heuristic algorithms with the optimal result derived by using CVX. Section \ref{conclusion} concludes the paper.

\section{Related Work}\label{sec:related_work}
Owing to the energy demands of DCs, many techniques and algorithms have been proposed to minimize the energy consumption of DCs \cite{Ghamkhari_2013}. Ghamkhari and Mohsenian-Rad \cite{Ghamkhari_2013} developed a mathematical model to capture the trade-off between the energy consumption of a data center and its revenue of offering Internet services. They proposed an algorithm to maximize the revenue of a DC by adapting the number of active servers according to the traffic profile. Fang \emph {et al.} \cite{Yonggang_Wen_Globemcom_2012} presented a novel power management strategy for the DCs, and their target was to minimize the energy consumption of switches in a DC. Cavdar and Alagoz \cite{Survey_Green_DC_2012} surveyed the energy consumption of servers and network devices of intra-DC networks, and showed that both computing resources and network elements should be designed with energy proportionality. In other words, it is better if the computing and networking devices can be designed with multiple sleeping states. A few green metrics are also provided by this survey, such as Power Usage Effectiveness (\emph{PUE}) and Carbon Usage Effectiveness~(\emph{CUE}).

Deng \emph {et al.} \cite{Fangming_Liu_IEEE_Network2014} presented five aspects of applying renewable energy in the DCs: the renewable energy generation model, the renewable energy prediction model, the planning of green DCs (i.e., various renewable options, avalabity of energy sources, different energy storage devices), the intra-DC work loads scheduling, and the inter-DC load balancing. They also discussed the research challenges of powering DCs with renewable energy.
Gattulli~\emph {et al.} \cite{IP_over_WDM_icc_2012} proposed algorithms to reduce $CO_{2}$ emissions in DCs by balancing the loads according to the renewable energy generation. These algorithms optimize renewable energy utilization while maintaining a relatively low blocking probability. Lin and Yu~\cite{Green_DC_CL_201704} proposed a distributed green networking approach to save energy for intra-DC networks by shutting off unused links. Kantarci \emph{et~al.}~\cite{inter_intra_DC_VM_placement} presented the intra-and-inter data center VM placement problem, and their objective was to minimize the energy consumption of all DCs and optical network components in the IP-over-WDM DC networks. They did not consider the renewable energy generation in this work. Meanwhile, EONs are different from WDM networks: they are more spectrum-efficient but more complicated, and scheduling requests in EONs are more complicated than WDM networks (because of the new constraints of EONs: path continuity constraint, spectrum continuity constraint and spectrum non-overlapping constraint). Fang \emph{et~al.}~\cite{Optimising_data_placement} investigated the power consumption problem of the network components in the backbone networks by jointly considering the data service placement and traffic flow routing.

Zhang \emph{et~al.}~\cite{Dynamic_service_placement} studied the dynamic service placement problem in a geographically distributed data center network; they proposed a framework to solve this problem based on the game theoretic models; they also considered a cloud platform shared by multiple service providers, and tried to minimize the operational cost of each service provider~\cite{Dynamic_service_placement}. Wu~\emph{et al.}~\cite{Renewable_DC_IP_WDM_2015} addressed the DC placement problem by jointly minimizing the brown energy consumption and the number of DCs, but they did not consider the constraint of the core network. Wu~\emph{et al.}~\cite{Green_DC_TGCN_2017} investigated the DC placement problem, the brown energy consumption problem, and DC cost problem in an inter-DC network with VM migration. Here, they assumed all requests are provisioned by the shortest path with unlimited network migration bandwidth. Mandal \emph{et~al.}~\cite{Green-cloud2013} studied renewable energy-aware VM migration techniques to reduce the brown energy consumption of DCs in an IP-over-WDM network. The VMs are uniformly placed across DCs, and the end-users are provisioned by the VMs through the shortest path in the beginning; then, the VM migration starts when the traffic load generation does not match the renewable energy generation: VMs in one DC, which lacks renewable energy, will be migrated to another DC, which has extra renewable energy. They proposed a migration-cost-aware algorithm to enhance the renewable energy utilization, and the VM migration happens when the migration cost (the cost of the brown energy consumption) is smaller than a pre-defined energy threshold; they also proposed a traffic-aware heuristic algorithm, which can relax the migration-cost-aware algorithm according to the traffic intensity. They considered path hops and the required bandwidth in the core network in migrating VMs among DCs. However, how to allocate bandwidth for the migration requests and how to assign a path in the optical network for the VM migration were not addressed.

Buyya \emph{et~al.} \cite{Buyya_utility_cloud_2010} investigated the architectural elements of the InterCloud for the utility-oriented federation of cloud computing environments. Ardagna \emph{et~al.} \cite{Ardagna_utility_DC_2012} studied the computing resource allocation problem in a multitier virtualized data center, and a linear utility function, which includes revenues and penalties, is used.
Ayoub.~\emph{et al.}~\cite{Efficient_VM_migration_JOCN_2017} investigated the routing and bandwidth assignment problem for an inter-DC network, and showed that the network performance can be improved by incorporating the path information in assigning bandwidth for VM migration. In our work, we assume the bandwidth required for migrating each VM is known \emph{a priori}; many VMs are aggregated together to form a data stream and then transmitted from one DC to anther DC via a lightpath; the same spectrum is used in the whole lightpath in EONs. VMs are taken to be independent from each other in this work; our work can be extended to multiple VMs belonging to the same application. If one application is provisioned by hosting VMs in multiple servers, we need to add two more variables to mark the set of servers selected for serving this application, and the set of VMs employed in the selected servers for this application. If we migrate one application from one DC to another DC, all VMs, which support this application, should be migrated.

EONs provide a higher connection capacity bandwidth, a lower spectrum granularity, a more flexible spectrum allocation, and more elasticity against time-varying traffic as compared to the WDM networks~\cite{EON_WDM_diff}. The large amount of traffic generated by the VM migration may congest the optical network even the optical network can provide huge bandwidth, and may increase the blocking rate of the network. EONs also incur more constraints as compared to the WDM networks in provisioning requests. Therefore, it is very challenging to investigate the RE-AIM problem in the inter-DC networks over the elastic optical infrastructure.

In our previous work~\cite{Liang_Cloudcom_2015} (Cloudcom paper), we studied the RE-AIM problem in an inter-DC network over the elastic optical infrastructure and presented preliminary results; we did not show how to map the VMs into lightpath requests nor did we show how to allocate network resources for user requests under a controllable backbone in migrating VMs. In this paper, we formulate the RE-AIM problem based on the ILP model; we use CVX and Gurobi to solve this ILP problem for small network instances and propose a few heuristic algorithms to solve the RE-AIM problem for large network configurations.

\section{Network Model and Problem Formulation}\label{sec:problem}
In this section, we present the network model, the energy model, and the formulation of the RE-AIM problem. The key notations are summarized in Table \ref{tab:notations}.

\begin{table}[!htb]
\scriptsize
\begin{center}
\caption{The Important Notations}\label{tab:notations}
\begin{tabular}{{|l|p{185pt}|}}
\hline
Symbol           & Definiton                                    \\
\hline
$c_{e}$          & The capacity of a link $e \in E$ in terms of spectrum slots.\\
$c_{m}$          & The maximum number of servers in the $m$th DC. \\
$c_{f}$          & The capacity of a frequency spectrum slot.         \\
$\phi$           & The maximum number of CPU cores in one server.               \\
$\alpha_{m}$     & Per unit energy cost for the $m$th DC.                    \\
$\beta_{m}$      & Per unit migration cost for the $m$th DC.                   \\
$\varsigma_{m}^{i}$ & The required bandwidth for the $i$th request in the $m$th DC. \\
$\psi_{m}^{i}$   & The required CPU cores for the $i$th request in the $m$th DC. \\
$\mathcal {R}$   & The traffic set.\\
$\kappa$         & The migration granularity defines the maximum bandwidth (capacity of a transmitter) that can be used in one migration.\\
$\upsilon_{p}$   & The used spectrum slot ratio of the $p$th path.\\
$\varGamma_{p, f}^{m, i}$& The number of used spectrum slots in the $p$th path and start from spectrum $f$ for the $i$th request from the $m$th DC.\\
$\upsilon_{max}$ & The maximum available network resources ratio.                 \\
$h_{max}$  &The maximum number of migrations allowed per DC.\\
$\mathcal {Q}_{h}$& The set of VM migration requests in the $h$th migration.\\
$P^{s}$          & The static power consumption of a server.                \\
$P^{d}$          & The dynamic power consumption of a server.                \\
$P^{i}$          & The power consumption of a server with no work load.       \\
$P^{p}$          & The power consumption of a server with full work load.      \\
$\eta$           & The power usage efficiency.                                  \\
$P_{m, n}$       & The power consumption of the $n$th server in the $m$th DC.    \\
$u_{m, n}$       & The CPU utilization of the $n$th server in the $m$th DC.       \\
$\xi_{m}$        & The amount of renewable energy generation in the $m$th DC.      \\
$\varPhi_{m}$    & The amount of brown energy consumption of the $m$th DC.          \\
$G$         & The bandwidth of a guard band for each transmission.         \\
$F_{max}$        & The upper bound of $f_{d, p}^{h}$, and it equals to the total bandwidth requirement of all requests in terms of spectrum slots,  $F_{max}=\sum \limits_{i=1}^{|\mathcal {R}|} \varsigma_{m}^{i} + G \cdot |\mathcal {R}|$.\\
\hline
\end{tabular}
\end{center}
\end{table}

\subsection{Network Model}\label{sec:network-model}
We model the inter-DC network by a graph, $\mathcal{G}(V, E, B)$. Here, $V$, $E$ and $B$ are the node set, the link set and the spectrum slot set, respectively. The set of DC nodes is denoted as $\mathcal{D}$. We assume that all DCs are powered by hybrid energy. We denote $\mathcal {D}_{s}$ as the set of DCs, which do not have sufficient renewable energy to support their work loads, and $\mathcal {D}_{d}$ as the set of DCs, which have surplus renewable energy. Each DC serves its own user requests in terms of provisioning VMs. Owing to the mismatch of the renewable energy and workload demands, we need to migrate VMs to alleviate this mismatch. During the migration, $\mathcal {D}_{s}$ and $\mathcal {D}_{d}$ refer to two different sets of DCs, acting as source and destination DCs for VM migration, respectively. We define $\kappa$ as the migration granularity, which determines the maximum routing resource that can be used in one migration to each DC.

\subsection{DC Model and Energy Model}\label{sec:DC-Energy-model}
Each request is provisioned by a VM, and different VMs may require different CPU resources. We assume all DCs have the same size, all servers in each DC have the same configuration, and the power consumption of each server includes static power consumption (fixed) and dynamic power consumption (variational).

Denote $P^{s}$, $P^{d}$, $P^{i}$ and $P^{p}$ as the static, dynamic, idle, and peak power consumption of a server, respectively. $P^{i}$ and $P^{p}$ define a server with no workload state and full workload state, respectively. $P_{m, n}$ is the power consumption of the $n$th server in the $m$th DC as shown in Eq.~\eqref{eq:em1}, and the CPU utilization of this server is defined as $u_{m, n}$ ($u_{m, n} \in [0, 1]$) \cite{Ghamkhari_2013}. The relationship among $P_{m, n}$, $P^{s}$, $P^{d}$ and $P^{p}$ are shown in  Eq.~\eqref{eq:em1}. Here, $\zeta$ is the number of used CPU cores of a server; $c_{s}$ is the maximum number of CPU cores in one server; $\eta$ is the PUE of a DC. PUE is defined as the total power consumption of a DC (including lighting, cooling, etc.~\cite{abbas2015_green_dc}) over all servers' power consumption in this DC. We also assume the idle servers are not turned off, and one server is in the idle state when it does not host any VM; otherwise, it is active. $\eta$ is set as $1.2$, $P^{i}$ as $100$ watts, and $P^{p}$ as $200$ watts according to \cite{Ghamkhari_2013}.

\begin{equation}\label{eq:em1}
\left\{
\begin{split}
& P_{m,n}=P^{s}+P^{d}     \\
& P^{s}=P^{i}+(\eta-1)P^{p}          \\
& P^{d}=(P^{p}-P^{i} )* u_{m, n}.    \\
& u_{m, n}=\zeta / c_{s}.
\end{split}
\right.
\end{equation}

Since the energy consumption of the core network equipment is very small as compared to that of the DC, we only consider the energy consumption of the DCs in this work. Denote $\xi_{m}$ as the amount of renewable energy available at the beginning of a migration cycle in the $m$th DC. Here, a migration cycle is the time interval, from the start time when the SDN controller refreshes the energy and workload status of all DCs for VM migration to the start time of the next update. Then, the brown energy consumption of the $m$th DC $\varPhi_{m}$ can be expressed in Eq. \eqref{eq:em3}.

\begin{equation}\label{eq:em3}
\varPhi_{m}=\max(\sum_{n} P_{m, n}-\xi_{m}, 0).
\end{equation}

\subsection{Problem Formulation}\label{sec:formulation}
The following variables are defined to formulate the RE-AIM problem.\\
$x_{m, n}^{i}=1$ if the $n$th server in the $m$th DC is used to provision the $i$th request; otherwise, it is $0$.\\
$y_{m, p}^{h}=1$ if the $p$th path is used in the $h$th migration from the $m$th DC; otherwise, it is~0.\\
$\omega_{m, n}^{d, i}=1$ if the $i$th request from the $d$th DC is migrated to the $n$th server of the $m$th DC; otherwise, it is $0$.\\
$z_{d}^{i}=1$ if the $i$th request in the $d$th DC is migrated out; otherwise, it is $0$.\\
$\theta_{d, m}^{h}$: a non-negative integer variable, which represents the required network bandwidth in the $h$th migration from the $d$th DC to the $m$th DC, and it is 0 if no migration happens.\\
$b_{d, p}^{h}$: a non-negative integer variable, which represents the required network bandwidth of the $p$th path in the $h$th migration from the $d$th DC, and it is 0 if no migration happens.\\
$f_{d, p}^{h}$: a non-negative integer variable, which represents the index of the starting spectrum slot in the $p$th path in the $h$th migration from the $d$th DC, and it is a positive integer if the $h$th migration from the $d$th DC is provisioned by using the $p$th path; otherwise, it is 0.

The objective of the RE-AIM problem is to minimize the total cost of the brown energy consumption of all DCs via VM migration by considering the DC service constraints, the transition constraints and the network constraints. Eq.~\eqref{eq:objective} is the objective function, where the first term represents the cost of the total brown energy consumption, the second term is the cost of VM migration in the DCs, and the third term is the cost of VM migration in the network. To emphasize the cost of the brown energy consumption, $\beta_{m}$ is relatively small as compared to $\alpha_{m}$. Here, Eqs.~\eqref{eq:e4}-\eqref{eq:e8} are DC service constraints, Eqs.~\eqref{eq:e9}-\eqref{eq:e14} are transition constraints, and the network constraints are shown in Eqs. \eqref{eq:e15}-\eqref{eq:e22}. The problem is formulated as follows.

Eq.~\eqref{eq:e4} constrains each request to be only provisioned once, i.e., one request can be provisioned locally or can be migrated to other DCs. Eq.~\eqref{eq:e5} constrains each request to be migrated no more than once. Eq.~\eqref{eq:e6} is the server CPU capacity constraint, and ensures the used CPU cores to be no more than the available ones. Eqs.~\eqref{eq:e7}-\eqref{eq:e8} impose the brown energy consumption constraint, which is transformed from Eq.~\eqref{eq:em3}.

Eq.~\eqref{eq:e9} is the VM migration bandwidth constraint, which ensures the used bandwidth from a DC is bigger than or equal to the bandwidth of all migrated VMs. Eqs.~\eqref{eq:e10}-\eqref{eq:e12} are path selection constraints, which ensure that only one path is selected for each migration. Eqs.~\eqref{eq:e10}-\eqref{eq:e11} ensure that the required network bandwidth for one migration is not zero when a path is used for migration. Eq.~\eqref{eq:e12} ensures that only one path is used for each migration. Here, $P(d, m)$ is the path set from the $d$th DC to the $m$th DC for all migrations, and $P(d, h)$ is the path set from the $d$th DC for the $h$th migration. Here, $F_{max}$ is the upper bound of $f_{d, p}^{h}$, which equals to the total bandwidth requirement in terms of spectrum slots. Eq.~\eqref{eq:e13} is the DC migration bandwidth constraint, which ensures that the required migration bandwidth from a DC equals to the total bandwidth of all used lightpaths of this DC. Eq.~\eqref{eq:e14} is the migration bandwidth capacity constraint, which ensures that the used bandwidth of a lightpath is smaller than the maximum bandwidth (capacity of the transceiver) in one transmission. Here, $L(\cdot)$ is the bit rate per symbol of a path which is determined by the modulation. For example, $L(p)=1$ if BPSK is used for the modulation.

\begin{align}
\label{eq:objective}
\min_{x,y,\omega,\theta,f}\quad & \sum_{m} ({\alpha_{m}\varPhi_{m} + \beta_{m}(\sum_{i}{z_{m}^{i} \cdot \varsigma_{m}^{i}}} +\sum_{h}\sum_{p} {y_{m,p}^{h}})  )\\
s.t.:\nonumber & \\
& DC\;service\;constraints:\nonumber\\
&\sum_{n}{x_{m, n}^{i}} + z_{d}^{i} \leq 1, \quad\forall i, d=m.                                      \label{eq:e4}\\
&\sum_{m}\sum_{n}\omega_{m, n}^{d, i}=z_{d}^{i}, \quad\forall i, d.                                   \label{eq:e5}\\
& \sum_{i} {x_{m, n}^{i} * \psi_{m}^{i}}+ \sum_{d}\sum_{i}\omega_{m, n}^{d, i} * \psi_{d}^{i} \leq \phi, \forall m, n. \label{eq:e6}\\
& \varPhi_{m} \geq \sum_{i} \sum_{n}{x_{m, n}^{i} * \psi_{m}^{i}} * P^{d} + n* P^{s} +  \nonumber\\
& \quad \sum_{d}\sum_{i}\sum_{n}  \omega_{m, n}^{d, i} * \psi_{d}^{i} * P^{d} - \xi_{m}, \quad\forall m.     \label{eq:e7}\\
& \varPhi_{m} \geq 0, \quad\forall m.                                                                            \label{eq:e8}\\
& Transition\;constraints:\nonumber\\
& \sum_{h}{\theta_{d,m}^{h}} \geq \sum_{i}\sum_{n} \omega_{m, n}^{d, i} * \varsigma_{d}^{i}, \quad\forall m, d. \label{eq:e9}\\
& \sum_{p \in P(d,m)}{y_{d,p}^{h}} * F_{max} \geq  \theta_{d,m}^{h}, \quad\forall m, d, h.             \label{eq:e10}\\
& \sum_{p \in P(d,m)}{y_{d,p}^{h}} \leq {\theta_{d,m}^{h}}, \quad\forall m, d, h.                           \label{eq:e11}\\
& \sum_{p \in P(d,h)}{y_{d,p}^{h}} \leq 1, \quad\forall d, h.                \label{eq:e12}\\
& \sum_{p \in P(d,m)}{b_{d,p}^{h}} = {\theta_{d,m}^{h}}, \quad\forall m, d, h.   \label{eq:e13}\\
& \sum_{m}{\theta_{d,m}^{h}} \leq \kappa* L(p), \quad\forall d, h.              \label{eq:e14}\\
& Network\;constraints:\nonumber\\
& \upsilon_{p}+ \frac{1}{c_{e}}*\sum_{d,h,f}\varGamma_{d, p}^{h, f}\leq \upsilon_{max}, \quad\forall p.        \label{eq:e15}\\
& y_{d, p}^{h} \leq f_{d, p}^{h} \leq y_{d, p}^{h} * F_{max}, \quad\forall d, p, h.\label{eq:e16}\\
& 0 \leq b_{d, p}^{h} \leq y_{d, p}^{h} * F_{max}, \quad\forall d, p, h.          \label{eq:e17}\\
&\begin{aligned}                                                                   \label{eq:e18}
& f_{d, p}^{h} +b_{d, p}^{h}-1+G \leq c_{e}*(\upsilon_{max}-\upsilon_{p}), \\
&\quad\quad\quad\quad\quad\quad\quad\quad\quad\quad\quad\quad \forall d, p, h.
  \end{aligned}
\end{align}

\begin{align}
& f_{d', p'}^{k}-f_{d, p}^{h} < \delta_{d,d'}^{h,k} * F_{max} \label{eq:e19}\\
& f_{d, p}^{h}-f_{d', p'}^{k} < (1-\delta_{d,d'}^{h,k}) * F_{max} \label{eq:e20}\\
& \begin{aligned}                                                         \label{eq:e21}
& f_{d, p}^{h}+b_{d, p}^{h}+G \cdot y_{d, p}^{h}-f_{d',p'}^{k}\leq F_{max} * \\
& \quad\quad  [2-\delta_{d,d'}^{h,k}- \gamma_{d,d'}^{h,k}], \quad\forall (d, h) \neq (d', k).
  \end{aligned}\\
& \begin{aligned}                                                         \label{eq:e22}
& f_{d', p'}^{k}+b_{d', p'}^{k}+G \cdot y_{d', p'}^{k} -f_{d, p}^{h}\leq F_{max} *  \\
& \quad\quad [1+\delta_{d,d'}^{h,k}-\gamma_{d,d'}^{h,k}], \quad\forall (d, h) \neq (d', k).
  \end{aligned}
\end{align}

Eq.~\eqref{eq:e15} constrains the network congestion ratio to be less than $\upsilon_{max}$, which is the maximum network congestion ratio allowed for routing in the network. In Eq. \eqref{eq:e15}, $\upsilon_{p}$ is the spectrum slot ratio of the $p$th path, which is defined as the ratio of the number of occupied spectrum slots in the $p$th path to the total number of spectrum slots of this path. $\varGamma_{d, p}^{h, f}$ is defined as the number of used spectrum slots in the $p$th path starting from the spectrum $f$ for the $h$th migration from the $d$th DC.

Eq.~\eqref{eq:e16} is the starting spectrum slot constraint, which ensures that the starting spectrum slot index of a path is a positive integer if this path is selected; it is zero if this path is not selected. Eq.~\eqref{eq:e17} is the spectrum allocation constraint, which ensures that the bandwidth of allocated spectrum to a path equals to the provisioned bandwidth. Here, $b_{m, p}^{h}$ is the required bandwidth of the $p$th path for the $h$th request from the $m$th DC, and $b_{m, p}^{h}$ equals to zero if no migration happens. Eq.~\eqref{eq:e18} is the available network bandwidth capacity constraint, which ensures that the bandwidth used in migrating VMs not to exceed the total available network resources.

\begin{align}
\delta_{d,d'}^{h,k}=
\begin{cases}
1, & f_{d, p}^{h} < f_{d', p'}^{k} \quad\quad\quad\quad\quad\quad\quad\\
0, & f_{d, p}^{h} \geq f_{d', p'}^{k} \quad\quad\quad\quad\quad\quad\quad
\end{cases}                         \label{eq:e24}  \\
\gamma_{d,d'}^{h,k}=
\begin{cases}
1, & path(d, h) \cap path(d', k)\neq \varnothing\\
0, & otherwise.
\end{cases}                        \label{eq:e25}
\end{align}

%

$\delta_{d, d'}^{h, k}$ is a Boolean variable as defined in Eq.~\eqref{eq:e24}, which equals $1$ if the starting spectrum slot index of the $p$th path in the $h$th migration from the $d$th DC is smaller than that of the $p'$th path for the $k$th migration from the $d'$th DC; otherwise, it is $0$. Since this definition is not linear, it is transformed into Eqs.~\eqref{eq:e19}-\eqref{eq:e20}. Eqs.~\eqref{eq:e21}-\eqref{eq:e22} are the spectrum non-overlapping and the continuity constraints. This spectrum non-overlapping constraint ensures that the spectrum used for two different paths does not overlap when the paths have more than one common link. Here, $\gamma_{d, d'}^{h, k}(\forall (d, h) \neq (d', k))$ is a Boolean indicator as defined in Eq.~\eqref{eq:e25}, which equals $1$ if the path used in the $h$th migration from the $d$th DC and that used in the $k$th migration from the $d'$th DC have at least one common link; otherwise, it is $0$. $path(\cdot)$ is a function to get the path information (all nodes in the path). We use an example to illustrate these equations. For example, if $f_{d, p}^{h} < f_{d', p'}^{k}$ and $path(d, h) \cap path(d', k)\neq \varnothing$, then Eq.~\eqref{eq:e20} ensures $\delta_{d, d'}^{h, k}=1$, Eq.~\eqref{eq:e19} is relaxed and $\gamma_{d, d'}^{h, k}=1$. Eq.~\eqref{eq:e21} becomes Eq.~\eqref{eq:e26}, which ensures the spectrum non-overlapping constraint and the continuity constraint. Eq.~\eqref{eq:e22} is automatically satisfied in this case. After that, Eq.~\eqref{eq:e26} is transformed to Eq.~\eqref{eq:e27} when $y_{d, p}^{h}=1$ ($f_{d, p}^{h}>0$), which ensures that the starting spectrum slot index of the $p'$th path in the $k$th migration from the $d'$th DC is bigger than the total occupied spectrum slots of the $p$th path in the $h$th migration from the $d$th DC and a guard band.


\begin{align}
f_{d, p}^{h}+b_{d, p}^{h}+G \cdot y_{d, p}^{h} \leq f_{d',p'}^{k} \label{eq:e26}\\
f_{d, p}^{h}+b_{d, p}^{h}+G \leq f_{d',p'}^{k}  \label{eq:e27}
\end{align}

In provisioning spectrum slots for requests in EONs, the path continuity constraint, spectrum continuity constraint and non-overlapping constraint must be considered.
For the path continuity constraint, a lightpath must use the same subcarriers in the whole path for a request. For the spectrum continuity constraint, the chosen subcarriers must be continuous if a request needs more than one subcarriers. For the non-overlapping constraint, two different lightpaths must be assigned with different subcarriers if they have one or more common links. Since we use a path based method to formulate the RE-AIM problem, the path continuity constraint of the network is already taken into account. Meanwhile, a guard band is required for each transmission.


\section{Problem Analysis and Heuristic Algorithms }\label{sec:analysis-algorithms}
\subsection{Problem Analysis}\label{sec:analysis}
The RE-AIM problem is tackled in three steps: i) determining VM migration requests in DCs, ii) performing routing and spectrum allocation (\emph{RSA}) in EONs, and iii) allocating computing resources in the DCs.
To solve the RE-AIM problem, both the energy costs in DCs and required network resources for the migration should be considered. For example, when a DC consumes brown energy, it is desirable to migrate some VMs to other DCs, a path should be allocated for routing from the source DC to the destination DC, and the spectrum slots should be assigned along this path to ensure the migration.  Therefore, it is challenging to solve RE-AIM, which is proven to be NP-hard by reducing any instance of the multiple knapsack problem into the RE-AIM problem, as detailed in the Appendix.


\begin{algorithm}[!htb]
\caption{Anycast with Shortest Path (\emph{Anycast-SP})}\label{Anycast-SP}
\SetKwData{Left}{left}\SetKwData{This}{this}\SetKwData{Up}{up}
\SetKwFunction{Union}{Union}\SetKwFunction{FindCompress}{FindCompress}
\SetKwInOut{Input}{Input}\SetKwInOut{Output}{Output}
\Input{$\mathcal{G}(V, E, B)$, $\mathcal{R}$, $\xi_{m}$, $\mathcal {D}$, and $\upsilon_{max}$\;}
\Output{$\mathcal {D}$, $\varPhi_{m}$ and $\mathcal {Q}_{h}$ \;}
\nl build $\mathcal {D}_{s}$ and $\mathcal {D}_{d}$ by the sub-optimal work loads allocation, and collect many-manycast requests\;
\nl set $block=0$, $h=0$\;
\nl \While{$\mathcal{D}_{s} \neq \varnothing \& \mathcal{D}_{d} \neq \varnothing \& block \neq 0$}{
\nl     h=h+1\;
\nl     find $s \in \mathcal {D}_{s}$ with the max migratory VMs\;
\nl     find $d \in \mathcal {D}_{d}$ with the max available renewable energy\;
\nl     sort the requests of the $s$th DC in ascending order by the bandwidth requirement\;
\nl     generate $\mathcal {Q}_{h}$ and get migration bandwidth $\theta_{d,m}^{h}$ which satisfies Eq.~\eqref{eq:e9} and \eqref{eq:e14}\;
\nl     calculate $\varGamma_{d,p}^{h,f}$ for the network congestion ratio\;
\nl     \If  {Eqs.~\eqref{eq:e6}, \eqref{eq:e10}-\eqref{eq:e13} $\&$ Eqs.~\eqref{eq:e15}-\eqref{eq:e22} are satisfied}{
\nl          path $y_{d,p}^{h}$ is used to migrate\;
\nl          find the start spectrum slot index $f_{d,p}^{h}$ in $B$\;
\nl          allocate the computing resource for $\mathcal {Q}_{h}$\;
\nl          update $\mathcal {D}_{s}$ and $\mathcal {D}_{d}$\;}
\nl    \Else{
\nl          $block=1$\; }
\nl     update all servers' status in $\mathcal {D}$\; }
\end{algorithm}

In the RE-AIM problem, the optimal workload distribution is calculated based on the current workload demands, green energy generation, and the availability of the network resources. The optimal solution is then derived based on the optimal workload distribution by using the CVX toolbox. Since it is difficult to obtain the optimal workload distribution, we use the sub-optimal workload distribution instead, which is calculated by relaxing the network constraints. Then, the heuristic algorithms execute the migration according to this sub-optimal workload distribution. For the RE-AIM problem, many VMs are migrated from many source DCs to many destination DCs, and it is a many-manycast communications problem. Since the RE-AIM problem is NP-hard, we propose a few heuristic algorithms to solve this problem by splitting many-manycast communications into many anycast communications. These algorithms determine which VM should be migrated to which DC and select a proper routing path in the network to avoid congesting the network, namely, Anycast with Shortest Path routing (\emph{Anycast-SP}) algorithm, Anycast with Maximum bandwidth Path routing (\emph{Anycast-MP}) algorithm, Anycast with Ergodic Path routing (\emph{Anycast-EP}) algorithm, and Anycast with Joint Resources Ergodic routing (\emph{Anycast-JRE}) algorithm, respectively.

\subsection{Heuristic Algorithms}\label{sec:H-Algorithms}
For all heuristic algorithms, the inputs are the traffic set $\mathcal{R}$, the available green energy $\xi_{m}$, the DC set $\mathcal {D}$, and $\upsilon_{max}$; the outputs are the DC set $\mathcal {D}_{s}$ which lack renewable energy, the DC set $\mathcal {D}_{d}$ which have abundant renewable energy, and the migration request set $\mathcal {Q}_{h}$ from $\mathcal {D}_{s}$ to $\mathcal {D}_{d}$ for each migration cycle.

\begin{algorithm}
\caption{Anycast with Maximum bandwidth Path (\emph{Anycast-MP})}\label{Anycast-MP}
\SetKwData{Left}{left}\SetKwData{This}{this}\SetKwData{Up}{up}
\SetKwFunction{Union}{Union}\SetKwFunction{FindCompress}{FindCompress}
\SetKwInOut{Input}{Input}\SetKwInOut{Output}{Output}
\Input{$\mathcal{G}(V, E, B)$, $\mathcal{R}$, $\xi_{m}$, $\mathcal {D}$, and $\upsilon_{max}$\;}
\Output{$\mathcal {D}$, $\varPhi_{m}$ and $\mathcal {Q}_{h}$ \;}
\nl build $\mathcal {D}_{s}$ and $\mathcal {D}_{d}$ by the sub-optimal work loads allocation\;
\nl collect many-manycast requests\;
\nl set $block=0$, $h=0$\;
\nl \While{$\mathcal{D}_{s} \neq \varnothing \& \mathcal{D}_{d} \neq \varnothing \& block \neq 0$}{
\nl     h=h+1\;
\nl     \For{all nodes $s \in \mathcal {D}_{s}$}{
\nl     \For {all nodes $d \in \mathcal {D}_{d}$}{
\nl           build K-shortest path set $\mathcal {P}$\;
\nl       \For  {path $p \in \mathcal {P}$}{
\nl            get the available bandwidth information for path $p$ under the current network configuration\;}}}
\nl     find path $p$ as the candidate path with the maximum available bandwidth\;
\nl     find $s$ and $d$ by the path information\;
\nl     get the migratory requests information in DC $s$, the server information in DC $d$, and get migration bandwidth $\theta_{d,m}^{h}$ which satisfies Eq.~\eqref{eq:e9} and \eqref{eq:e14}\;
\nl     steps~ $11$--$18$ of Alg. \ref{Anycast-SP}\;}
\end{algorithm}

The Anycast-SP algorithm, as shown in Alg. \ref{Anycast-SP}, is to find the shortest routing path that satisfies the VM migration requirement and the network resource constraints.
The migration will try to use the shortest path $p$ from $s$ to $d$; the request set $\mathcal {Q}_{h}$ is carried out if the network congestion constraint is satisfied; otherwise, the migration is denied.  Afterward, we update $\mathcal {D}_{s}$ and $\mathcal {D}_{d}$ for the next migration. After many rounds of migration, if $\mathcal {D}_{s}$ or $\mathcal {D}_{d}$ is empty, or Eq. (\ref{eq:e9}) is not satisfied, the migration is terminated. Details of the Anycast-SPR algorithm is described in \emph{Algorithm} \ref{Anycast-SP}.

The complexity of Anycast-SP is $O(|B| |E|^{2}\mathcal{|R|}\mathcal|{Q}_{h}|+\mathcal {|D|}^{2}c_{m}\phi)$. Here, $O(\mathcal {|D|}^{2}c_{m}\phi)$ is the complexity to determine the sub-optimal work loads, $O(|B|)$ is the complexity to determine the starting spectrum slot index, and $O(\mathcal{|R|}\mathcal|{Q}_{h}|)$ is the complexity in building the VM set for the migration. $O(|E|^{2})$ is the complexity of determining the path for Anycast-SP.

\begin{algorithm}
\caption{Anycast with Joint Resources Ergodic routing (\emph{Anycast-JRE})}\label{Anycast-JRE}
\SetKwData{Left}{left}\SetKwData{This}{this}\SetKwData{Up}{up}
\SetKwFunction{Union}{Union}\SetKwFunction{FindCompress}{FindCompress}
\SetKwInOut{Input}{Input}\SetKwInOut{Output}{Output}
\Input{$\mathcal{G}(V, E, B)$, $\mathcal{R}$, $\xi_{m}$, $\mathcal {D}$, and $\upsilon_{max}$\;}
\Output{$\mathcal {D}$, $\varPhi_{m}$ and $\mathcal {Q}_{h}$ \;}
\nl build $\mathcal {D}_{s}$ and $\mathcal {D}_{d}$ by the sub-optimal work loads allocation\;
\nl collect many-manycast requests\;
\nl set $h=0$\;
\nl \While{$\mathcal{D}_{s} \neq \varnothing \& \mathcal{D}_{d} \neq \varnothing$}{
\nl     h=h+1\;
\nl     \For{all nodes $s \in \mathcal {D}_{s}$}{
\nl     \For {all nodes $d \in \mathcal {D}_{d}$}{
\nl           build K-shortest path set $\mathcal {P}$\;
\nl           calculate the available computing resource information for the DC $d$\;
\nl       \For  {path $p \in \mathcal {P}$}{
\nl            get the weight of path $p$ by Eq.~\eqref{eq:w1} under the current network configuration\;}}}
\nl     find the path $p$ as the candidate path with the maximum weight\;
\nl     find $s$ and $d$ by the path information\;
\nl     get the migratory requests information in the DC $s$, and the server information in the DC $d$\;
\nl     sort the requests of the $s$th DC in an ascending order by the bandwidth requirement\;
\nl     use the sorted requests to generate migration request set $\mathcal {Q}_{h}$ for the $h$th migration, and get migration bandwidth $\theta_{d,m}^{h}$ which satisfies Eq.~\eqref{eq:e9} and \eqref{eq:e14}\;
\nl     calculate $\varGamma_{d,p}^{h,f}$ for the network congestion ratio\;
\nl     \If  {Eqs.~\eqref{eq:e6}, \eqref{eq:e10}-\eqref{eq:e13} $\&$ Eqs.~\eqref{eq:e15}-\eqref{eq:e22} are satisfied}{
\nl          path $y_{d,p}^{h}$ is used to migrate\;
\nl          find the start spectrum slot index $f_{d,p}^{h}$ in $B$ \;
\nl          allocate the computing resource for $\mathcal {Q}_{h}$ in the DC $d$\;}
\nl    \Else{
\nl      add one DC node ($s$ or $d$) with less energy migration requirement to $\mathcal {D}_{out}$\;}
\nl     update $\mathcal {D}_{s}$ and $\mathcal {D}_{d}$ by removing nodes in $\mathcal {D}_{out}$\;
\nl     update all servers' status in $\mathcal {D}$\; }
\end{algorithm}

When the work load of the network is heavy or too many VMs need to be migrated, it is difficult for the Anycast-SP algorithm to find the shortest path with available spectrum slots. Then, Anycast-SP may block some migration requests, and leads to high brown energy consumption of DCs. Here, we propose another benchmark algorithm (Anycast-MP), which places more weight on network resources. Anycast-MP checks $K$-shortest paths from the source node to the destination node, and picks up the idlest path to provision the migration requests. It aims to find a path with more available spectrum slots at the expense of a higher complexity. The main difference between Anycast-MP and Anycast-SP is using different ways to determine a path. Details of the Anycast-MP algorithm is described in \emph{Algorithm} \ref{Anycast-MP}. The complexity of Anycast-MP is $O(K |B| |E|^{2}\mathcal{|R|}\mathcal|{Q}_{h}|+\mathcal{|D|}^{2} c_{m}\phi)$. Here, $O(\mathcal {|D|}^{2}c_{m}\phi)$ is the complexity to determine the sub-optimal work loads, $O(|B|)$ that to determine the starting spectrum slot index, and $O(\mathcal{|R|}\mathcal|{Q}_{h}|)$ that to build the VM set for the migration. $O(K|E|^{2})$ is the complexity of determining the path for Anycast-MP. The most complex part is to determine the set of VMs for the migration.

To better address the RE-AIM problem, two Ergodic routing algorithms, Anycast-JRE and Anycast-EP, are proposed. Anycast-JRE checks $K$-shortest paths from the source node to the destination node, also checks the available computing resources of the destination DCs, and picks up the path with the maximum weight to provision the migration requests. One DC with less energy migration requirement will be added to $\mathcal {D}_{out}$ if this migration fails, and this DC will be removed from $\mathcal {D}_{s}$ and $\mathcal {D}_{d}$. The migration will continue until $\mathcal {D}_{s}$ or $\mathcal {D}_{d}$ is empty. Details of the Anycast-JRE algorithm is described in \emph{Algorithm} \ref{Anycast-JRE}. The complexity for Anycast-JRE is $O(K^{2} |B|^{2} |E|^{2} \mathcal{|R|} \mathcal|{Q}_{h}|\mathcal{|D|}c_{m}\phi+ \mathcal{|D|}^{2}c_{m}\phi)$. Here, $O(\mathcal {|D|}^{2}c_{m}\phi)$ is the complexity to determine the sub-optimal work loads, $O(|B|)$ that to determine the starting spectrum slot index, and $O(\mathcal{|R|}\mathcal|{Q}_{h}|)$ that to build the VM set for the migration. $O(K|B|\mathcal{|D|}c_{m}\phi)$ is the complexity of calculating the weight $W(p,d)$. $O(K|E|^{2})$ is the complexity of determining path for Anycast-JRE. The most complex parts are to determine the set of VMs for the migration and to calculate the weight.

\begin{equation}\label{eq:w1}
W(p,d)=(A(p)/H(p))\cdot \sum_{n}{(1-u_{m, n})*\phi}, \quad m=d
\end{equation}
\begin{equation}\label{eq:w2}
 W(p,d)=A(p)/H(p), \quad m=d
\end{equation}

Anycast-JRE with Ergodic Path routing becomes Anycast-EP if Eq.~\eqref{eq:w1} is replaced by Eq.~\eqref{eq:w2} to calculate the weight, and step $9$ is deleted. The complexity for Anycast-EP is $O(K^{2} |B|^{2} |E|^{2} \mathcal{|R|} \mathcal|{Q}_{h}| + \mathcal{|D|}^{2} c_{m}\phi)$, which is the same as Anycast-JRE except that the complexity of calculating weight $W(p,d)$ is $O(K|B|)$. The most complex part is to determine the set of VMs for the migration.



\subsection{Algorithm Implementation}\label{sec:Implementation}
Here, we show how to implement the algorithms in the inter-DC network. The core optical network is a centralized architecture and optical circuit switching (\emph{OCS}) is employed for swapping data. There is a controller in this network, and our algorithms are running inside the controller. For each migration cycle, each DC will exchange its green energy information, work load information, server information and network utilization information from the controller. It takes a very short time (less than one second or a few seconds) for the heuristic algorithms to execute the migration. Here, one migration cycle is set as one hour in our simulation. When the controller receives the exchanged information from all DCs, it runs one the four proposed heuristic algorithms to calculate VM migration requests. After that, these VMs are migrated among DCs through the EON. The whole process is repeated until it is terminated by the controller.

\section{Performance Evaluations}\label{sec:evaluations}
The algorithms for solving the RE-AIM problem are evaluated in this section. MATLAB is used for the simulations, which are run on a Dell desktop with $4$-core $8$-line Intel Core $i7-3770$ and $16$ GB RAM. In this work, we assume VM migration is carried out in every migration cycle, and a migration cycle is set as one hour. Meanwhile, the user requests provisioned by VMs (last for one hour or more). For the VMs which run for more than one hour, they may be migrated in every migration cycle, depending on the workload distribution and the availability of the renewable energy. We use CVX \cite{CVX} with Gurobi \cite{Gurobi} to solve the ILP problem.

\begin{table}
\centering
\scriptsize
\caption{\small Results for the Six-Node Topology }\label{tab:ILP-six-node}
\begin{tabular}{|p{40pt}|c|c|c|c|c|c|}
\hline
     & \multicolumn{3}{|c|}{2 requests per DC}  & \multicolumn{3}{|c|}{3 requests per DC}  \\
      & \multicolumn{3}{|c|}{$h_{max}=1$}  & \multicolumn{3}{|c|}{$h_{max}=2$}  \\
\cline{2-7}
Algorithms                &   Obj & Obj2 & Time  & Obj &Obj2 & Time   \\
\hline
Without\quad Migration    & 1849.5 &1849.5& 0         & 1742.6& 1742.6  & 0  \\
\hline
ILP                       & 1169.9 &1168.7&224.29     & 1160  & 1158.9  &2035 \\
\hline
Anycast-SP                & 1484.8&1484.1 &0.02       &  1501.9& 1501.2 &0.02   \\
\hline
Anycast-MP                & 1432.8&1432.1  &0.04      &  1444.1& 1443.4 & 0.04  \\
\hline
Anycast-EP                & 1216.3&1215.3&0.07        &  1311.1& 1310.1 &0.09    \\
\hline
Anycast-JRE               &1213   &1211.8 &0.09       &  1186.4& 1185.2 &0.09   \\
\hline
\end{tabular}
\end{table}

\begin{figure}[!htb]
    \centering
    \includegraphics[width=1.0\columnwidth]{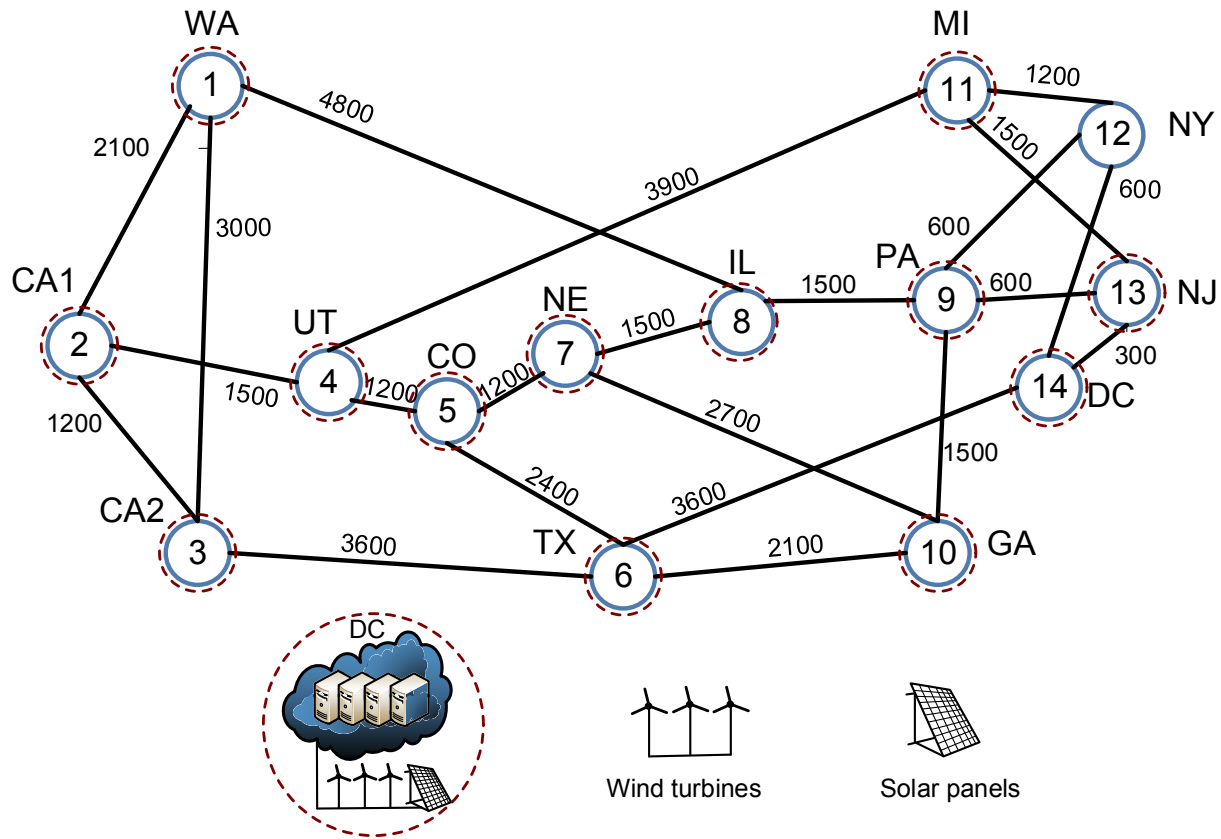}
    \caption{\small NSFNET topology (link length in km).}
    \label{fig:NSF-green}
\end{figure}

\subsection{Evaluations of small scale problems}\label{sec:small-evaluations}
We evaluate the performance of the heuristic algorithms, and compare the results with the optimal results derived by using CVX for small network configurations with $2$ user requests per DC for the six-node topology, $3$ user requests per DC for the six-node topology, and $2$ user requests per DC for the NSF topology. For all simulations, we repeat each simulation five times to obtain the average value. Tables~\ref{tab:ILP-six-node} and \ref{tab:ILP-NSF} summarize the simulation results for the six-node topology (the same as Fig.~\ref{fig:six-cloud} with 1200 $km$ length for each link) and NSF topology (Fig.~\ref{fig:NSF-green}), respectively. The user requests for each DC is generated according to a Poisson distribution. There are $6$ and $14$ DCs for the six-node topology and NSF topology, respectively. $K$ is set to $3$ and $1$ for the six-node topology and NSF topology, respectively. $c_{m}$ is set to $1$ server, and the price of electricity $\alpha$ is randomly chosen in a range $[9, 15]$ cents. The other parameters used for our evaluations are the same as Table~\ref{tab:simulation-parameters}.

\begin{table}
\centering
\scriptsize
\caption{\small Results for the NSF Topology }\label{tab:ILP-NSF}
\begin{tabular}{|p{70pt}|c|c|c|}
\hline
     & \multicolumn{3}{|c|}{2 requests per node, $h_{max}=1$}   \\
\cline{2-4}
Algorithms                &   Obj & Obj2  & Time  (sec)\\
\hline
Without Migration         & 2427.9&2427.9&0         \\
\hline
ILP                       &1275&1273.4    &1533.5   \\
\hline
Anycast-SP                &1844.7 &1844   &0.02      \\
\hline
Anycast-MP                &2269.5 &2269.3 &0.03      \\
\hline
Anycast-EP                &1342.7 &1341.1 &0.09      \\
\hline
Anycast-JRE               &1321.6 &1320   &0.07      \\
\hline
\end{tabular}
\end{table}

\begin{table}[!htb]
\centering
\scriptsize
\begin{center}
\caption{\small Simulation Parameters}\label{tab:simulation-parameters}
\begin{tabular}{{|l|p{150pt}|}}
\hline
Network topology                    & NSFNET\\
\hline
$\mathcal {D}$                      & \{1, 2, 3, ..., 14\}             \\
\hline
$\phi$                              & $16$ CPU cores                        \\
\hline
$c_{m}$                             & $100$ servers                     \\
\hline
$\alpha=\cup \{\alpha_{m}\}$        &  $\{9.09, 11.28, 12.57, 10.88, 12.12, 11.56$ \\
                                    &  $10.60, 12.50, 13.64, 11.54, 14.42, 18.54,$  \\
                                    &  $15.81, 12.99 \}$ cents                         \\
\hline
$\xi_{m}$                           & $[0.3*P^{p}c_{m}\eta, P^{p}c_{m}\eta]$ \\
\hline
$\beta_{m}$                         &  $\$0.001$ per unit                      \\
\hline
$P^{i}$                             & $100$ W                               \\
\hline
$P^{p}$                             & $200$ W                                 \\
\hline
$\eta$                              & $1.2$                                  \\
\hline
$c_{e}$                             & $300$ spectrum slots                       \\
\hline
$c_{f}$                             & $12.5$ Gbps                                  \\
\hline
$K$                                 & $3$ paths                                    \\
\hline
$\kappa$                            & 100 Gbps                                      \\
\hline
$\psi_{m}^{i}$                      & $[1, 3]$ CPU cores                               \\
\hline
$\varsigma_{m}^{i}$                 & $[2, 20]$ Gbps                        \\
\hline
requests per node                   & $\{400, 440, 480, 520, 560, 600, 640, 680\}$      \\
\hline
\end{tabular}
\end{center}
\end{table}

Here, ``obj" is referred to the objective defined in Eq.~\eqref{eq:objective}, and ``obj2" is referred to the total cost of brown energy consumption ($obj2=\sum_{m} {\alpha_{m}\varPhi_{m}}$). Table \ref{tab:ILP-six-node} shows the results of all heuristic algorithms and the optimal results derived by using CVX in the six-node topology. The number of migrations are set to $h_{max}=1$ and $h_{max}=2$, respectively. The performance of Anycast-EP and Anycast-JRE are very close to the optimal results. Anycast-MP's performance is a litter better than that of Anycast-SP. Even there are only $2$ user requests per DC, it takes nearly four minutes to achieve the optimal result, and it takes nearly 34 minutes to achieve the optimal result for $3$ user requests per DC.

Table \ref{tab:ILP-NSF} shows the the results of all heuristic algorithms and the optimal result derived by using CVX for the NSF topology. The gap between the optimal result and that of Anycast-EP is $3.6\%$, and the gap between the optimal result and that of Anycast-JRE is $5.3\%$. Anycast-MP's performance is a litter worse than that of Anycast-SP, because Anycast-MP places more weight on the bandwidth of the network, and the destination DC selected by Anycast-MP may not have computing resources.

\subsection{Evaluations of large scale problem}\label{sec:large-evaluations}
The NSFNET topology~\cite{liang_globecom_2016, zhang-osa} consists of $14$ nodes with DCs are located at $\mathcal {D}= \{1, 2, 3, ..., 14\}$. The DCs are assumed to be equipped with wind turbines and solar panels, which provide the DCs with renewable energy. The constant $\alpha$ is set as the price of electricity~\cite{electricity_price2016}. The capacity of a spectrum slot $c_{f}$ is set to 12.5Gbps. The capacity of the network $c_{e}$ is set to $300$ spectrum slots, and $\upsilon_{max}c_{e}$ is the number of available spectrum slots for migration. $K$, i.e., the maximum number of shortest paths that can be used in Anycast-JRE (EP), is $3$. The VM bandwidth requirement $\varsigma_{m}^{i}$ is randomly selected from $[2, 20]$ Gbps, and the computing requirement $\psi_{m}^{i}$ is randomly selected from $[1, 3]$ CPU cores. Parameters which are used for the evaluation are summarized in Table~\ref{tab:simulation-parameters}.

\begin{figure}[!htb]
    \centering
    \includegraphics[width=1.0\columnwidth]{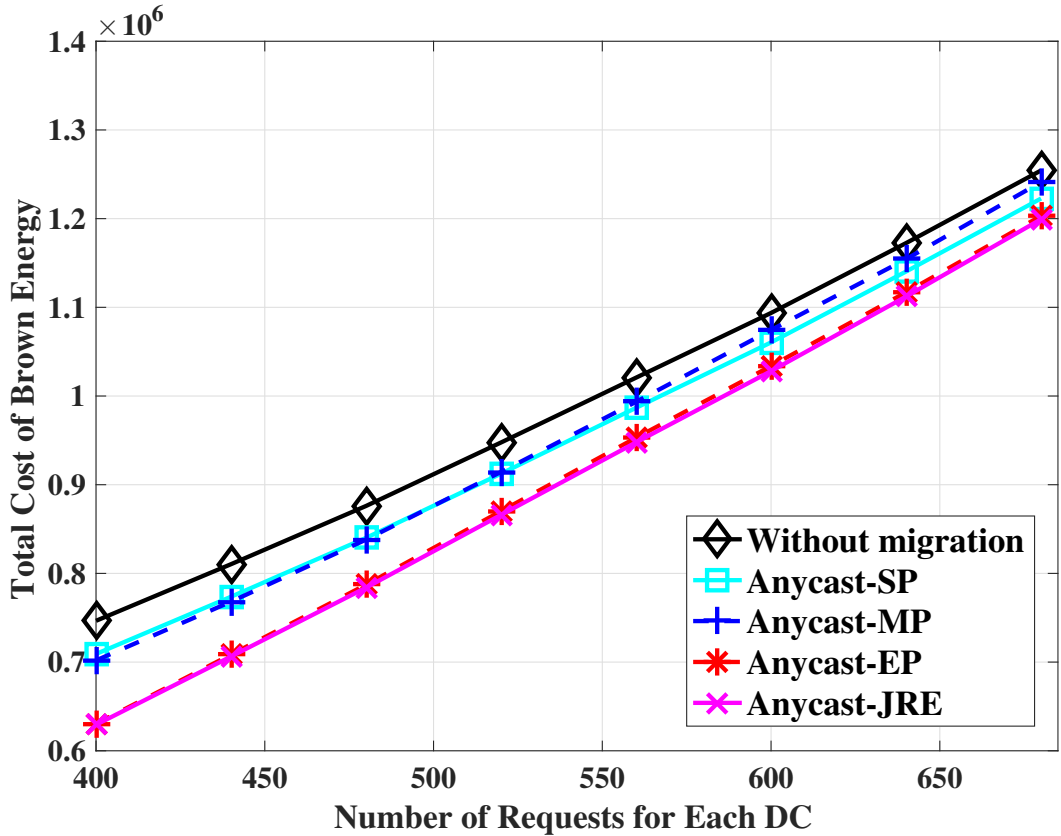}
    \caption{\small Total cost of brown energy consumption under $\upsilon_{max}=0.5$.}
    \label{fig:Energy-0.5}
\end{figure}

\begin{figure}[!htb]
    \centering
    \includegraphics[width=1.0\columnwidth]{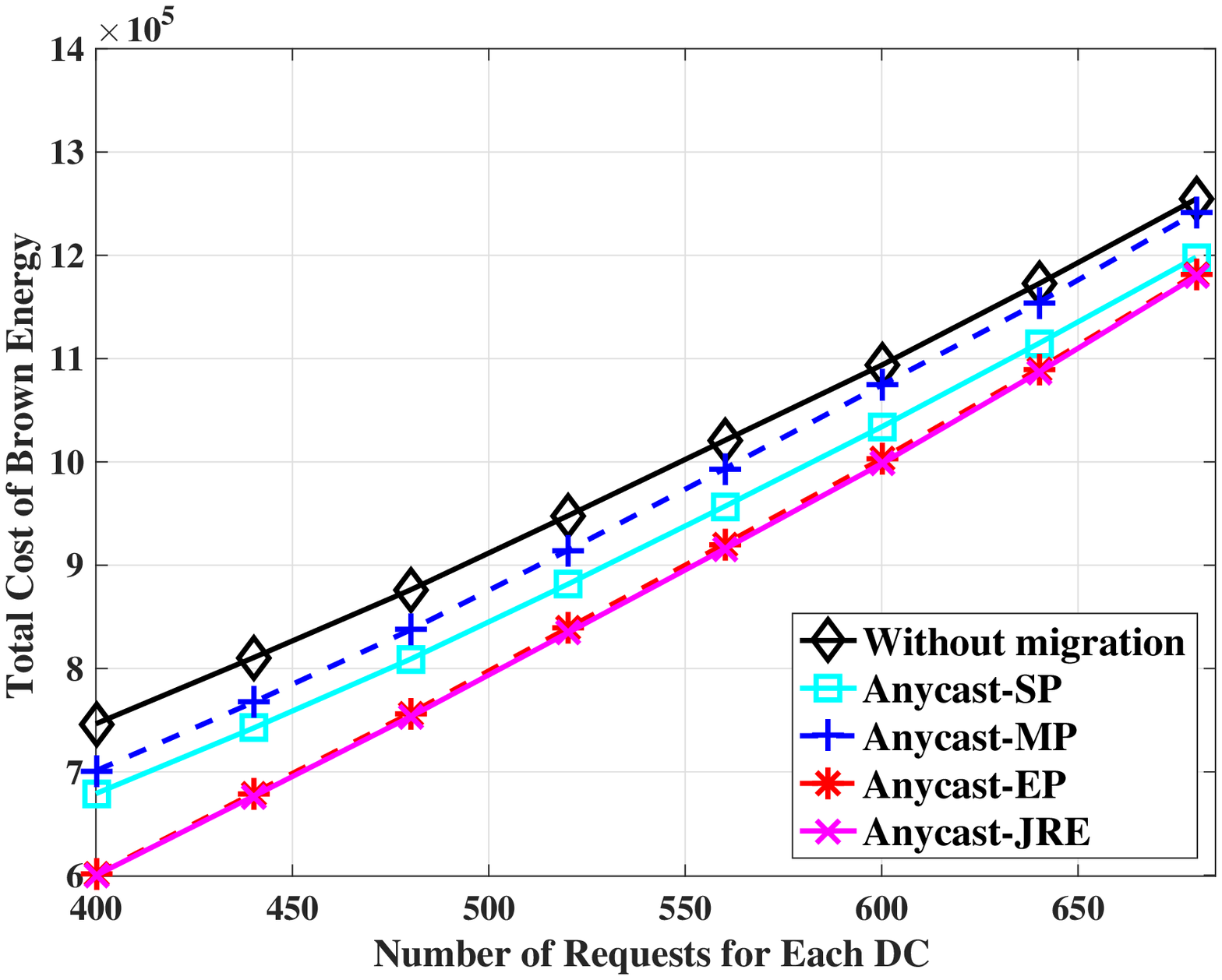}
    \caption{\small Total cost of brown energy consumption under $\upsilon_{max}=1.0$.}
    \label{fig:Energy-1.0}
\end{figure}

We repeat the simulation $200$ times. Fig.~\ref{fig:Energy-0.5} shows the total cost of brown energy consumption of different provisioning strategies when $\upsilon_{max}$ equals to $0.5$. Apparently, all algorithms can save brown energy substantially as compared to without migration. Anycast-SP, Anycast-MP, Anycas-EP, and Anycas-JRE save up to $5.0\%$, $6.0\%$, $15.6\%$, and $15.7\%$ cost of brown energy as compared with the strategy without migration, respectively. Anycast-JRE and Anycast-EP achieve better performance because they check for more possible paths and DCs, while Anycast-SP and Anycast-MP stop immediately when one migration fails. Fig.~\ref{fig:Energy-1.0} shows the total cost of brown energy consumption of different provisioning strategies when $\upsilon_{max}$ equals to $1.0$. Anycast-SP, Anycast-MP, Anycas-EP, and Anycas-JRE save up to $9.0\%$, $6.1\%$, $19.5\%$, and $19.7\%$ cost of brown energy as compared with the strategy without migration, respectively. Nearly all algorithms achieve better performance for $\upsilon_{max}=1.0$  as compared to $\upsilon_{max}=0.5$. Anycast-MP always uses the path with the largest bandwidth to provision the VM migration requests, and thus can accommodate the least number of VM migrations.

Figs.~\ref{fig:Energy-0.5}-\ref{fig:Energy-1.0} show that the total cost of the brown energy consumption increases as the workload increases; migration cannot reduce the cost of the brown energy consumption much under a heavy work load as compared to the strategy without migration because the renewable energy of all DCs is nearly fully utilized by their own work loads. However, this does not mean that migration is useless, but rather that less migration is needed for heavy work loads.

\begin{figure}[!htb]
    \centering
    \includegraphics[width=1.0\columnwidth]{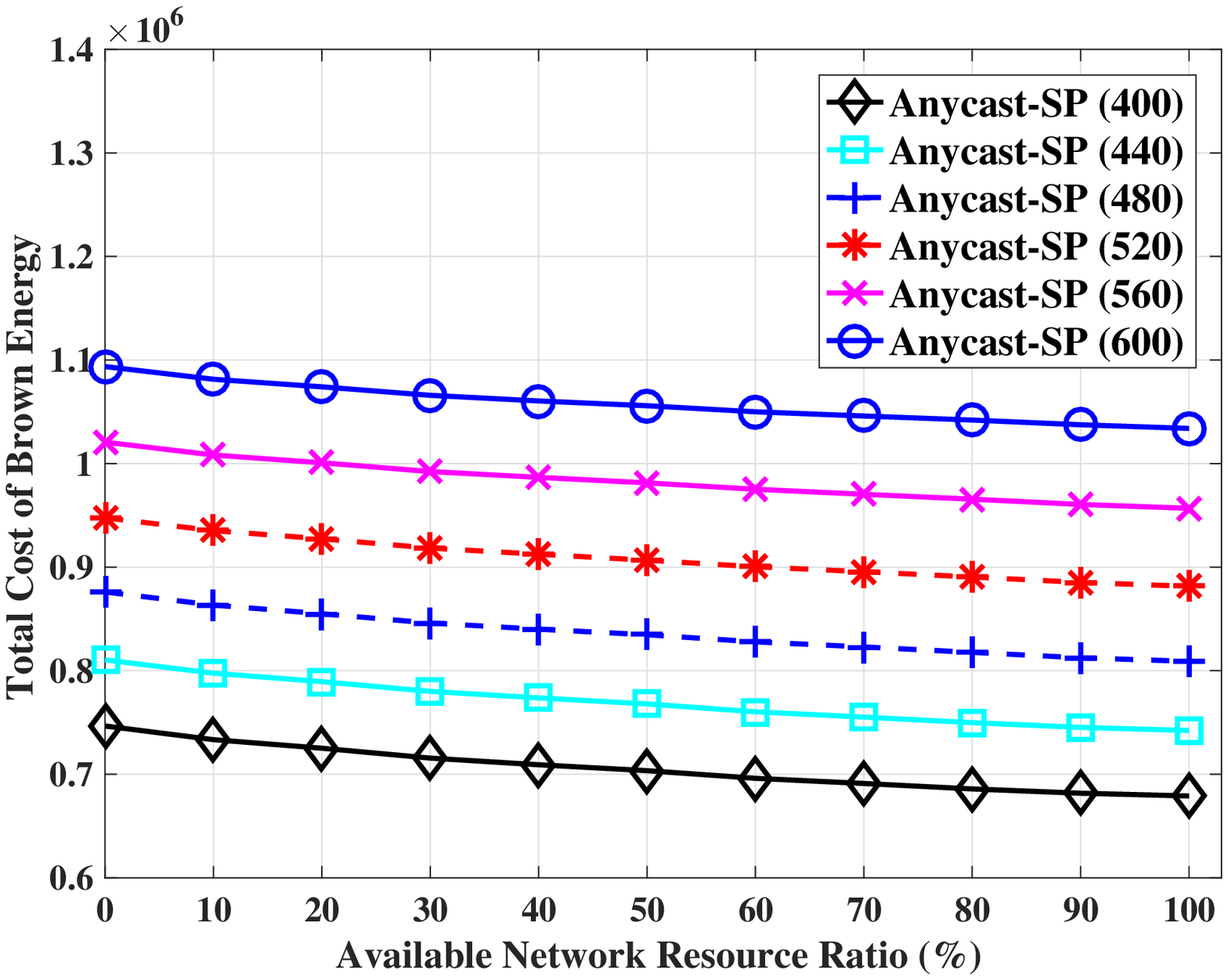}
    \caption{\small Anycast-SP performance under different $\upsilon_{max}$.}
    \label{fig:SP_result}
\end{figure}

\begin{figure}[!htb]
    \centering
    \includegraphics[width=1.0\columnwidth]{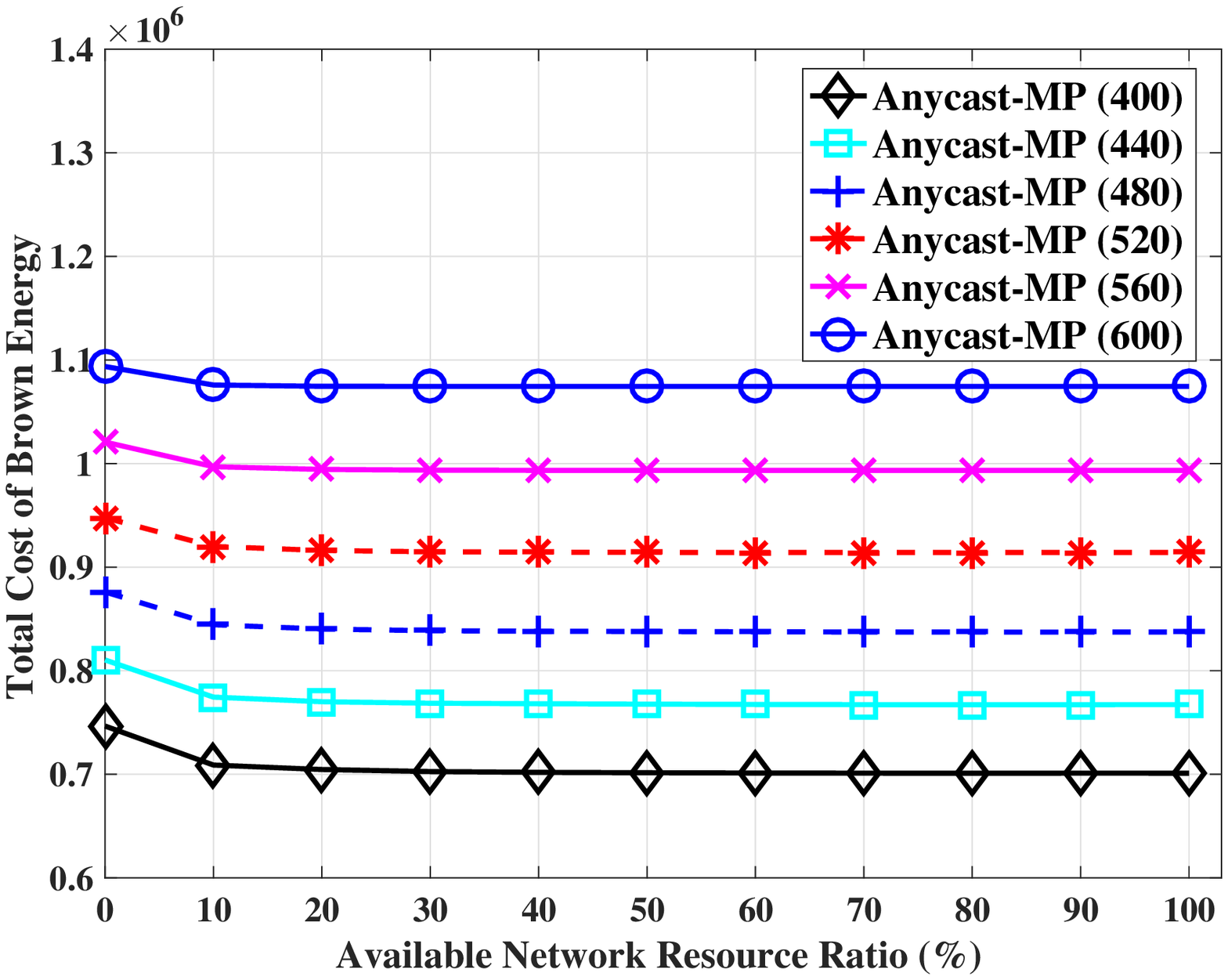}
    \caption{\small Anycast-MP performance under different $\upsilon_{max}$.}
    \label{fig:MB_result}
\end{figure}

\begin{figure}[!htb]
    \centering
    \includegraphics[width=1.0\columnwidth]{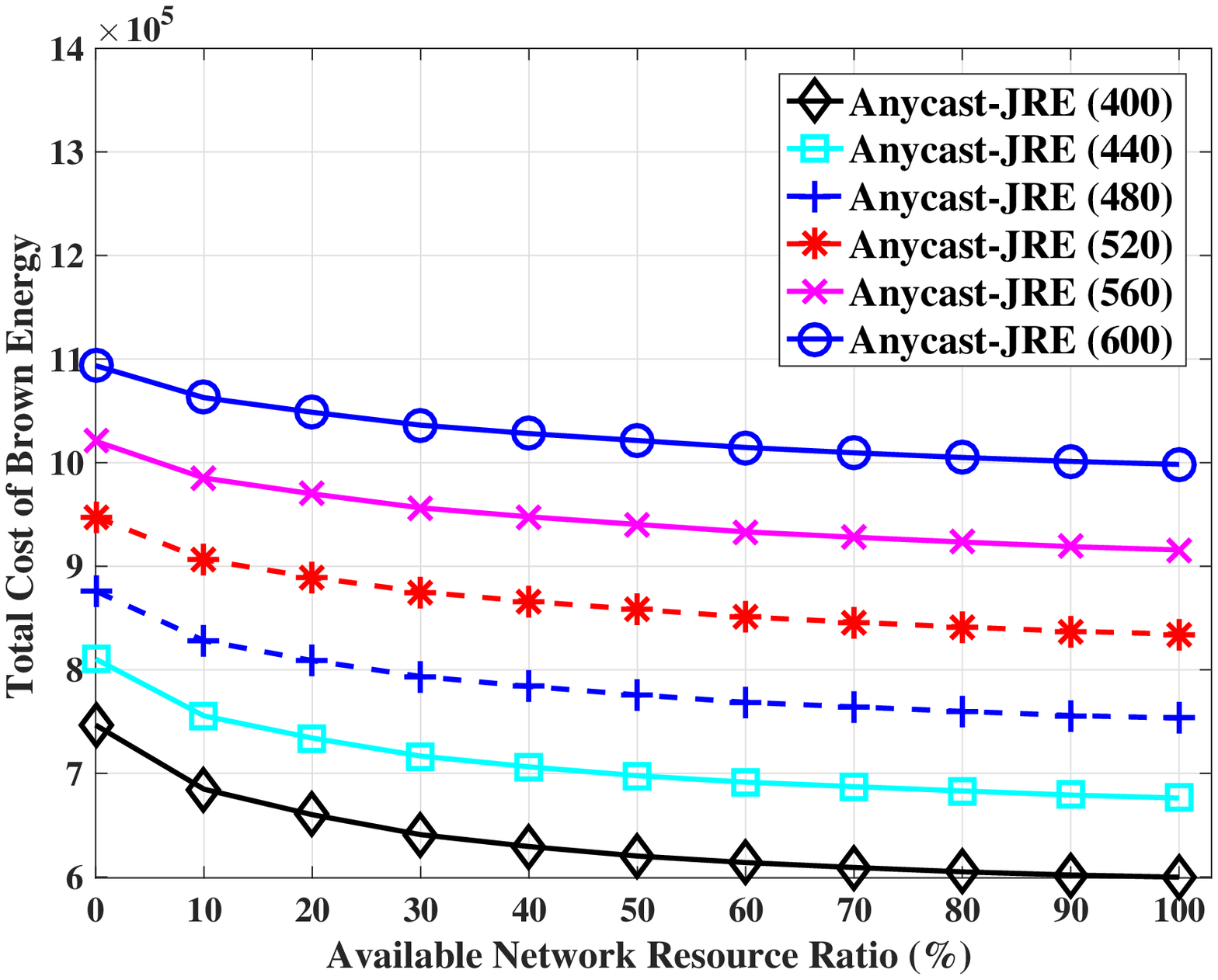}
    \caption{\small Anycast-JRE performance under different $\upsilon_{max}$.}
    \label{fig:JEB_result}
\end{figure}


\begin{figure}[!htb]
    \centering
    \includegraphics[width=1.0\columnwidth]{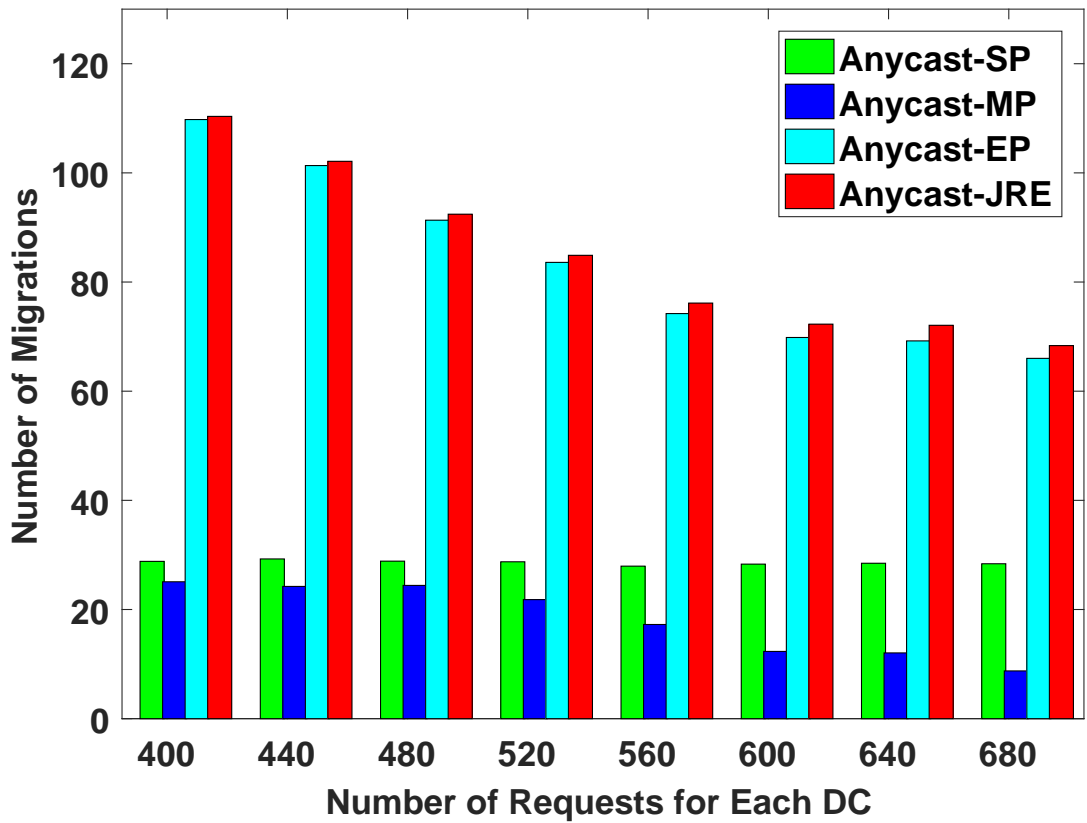}\caption{\small Number of migrations under $\upsilon_{max}=0.5$.}
    \label{fig:migration-0.5}
\end{figure}

\begin{figure}[!htb]
    \centering
    \includegraphics[width=1.0\columnwidth]{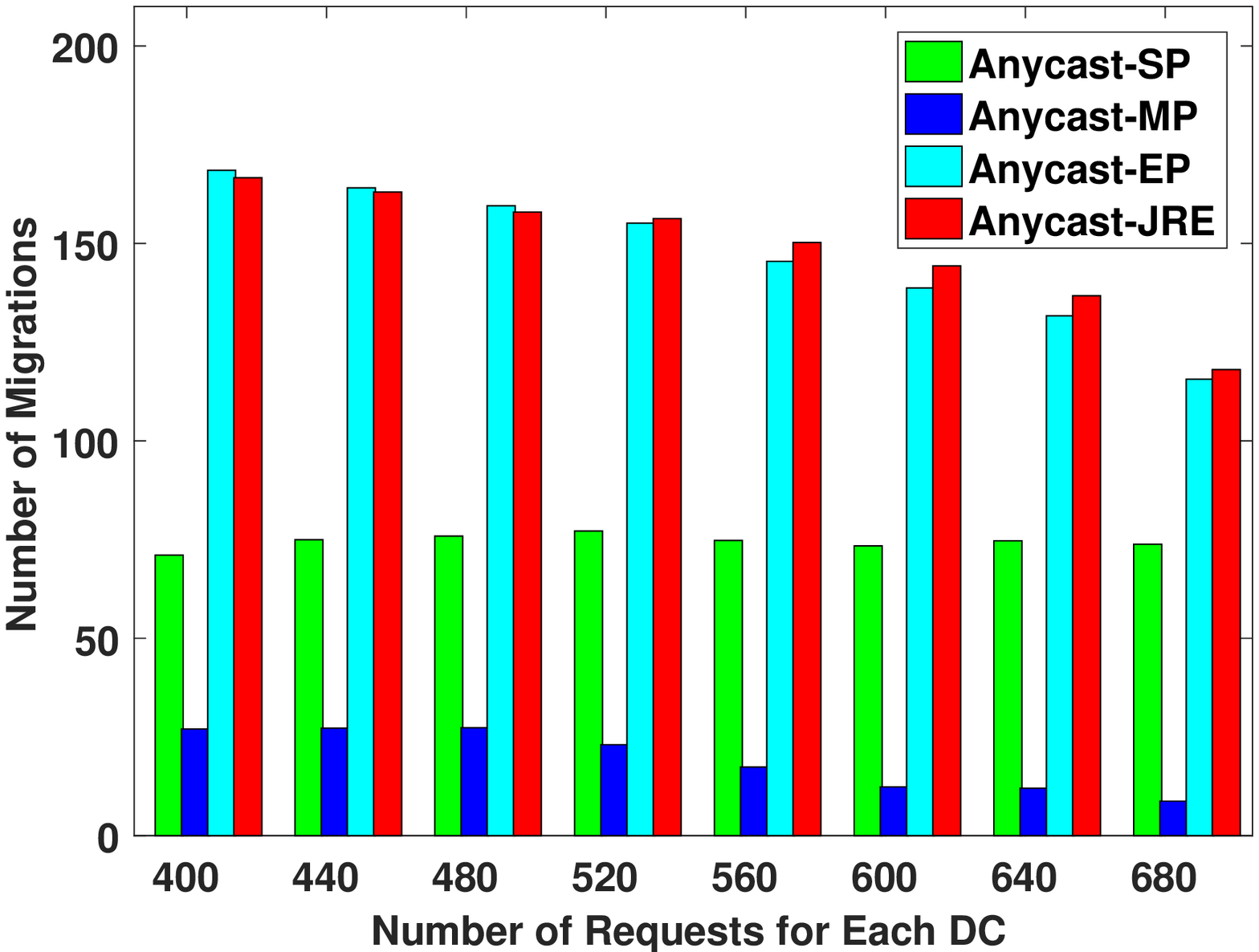}\caption{\small Number of migrations under $\upsilon_{max}=1.0$.} \label{fig:migration-1.0}
\end{figure}

In order to obtain a better analysis, we evaluate the algorithms under different $\upsilon_{max}$ from $0$ to $100\%$, as shown in Figs.~\ref{fig:SP_result}--\ref{fig:JEB_result}. All algorithms show that the total cost of brown energy consumption increases as traffic load increases for a fixed $\upsilon_{max}$. The results also show that the total cost of brown energy consumption decreases as $\upsilon_{max}$ increases, when the traffic load is fixed because more network bandwidth resource is available for VM migration for a bigger $\upsilon_{max}$. Anycast-JRE incurs the highest computing complexity and hence achieves the lowest cost of the brown energy consumption.

Figs.~\ref{fig:migration-0.5}--\ref{fig:migration-1.0} show the number of migrations under $\upsilon_{max}=0.5$ and $\upsilon_{max}=1.0$, respectively. Anycast-MP incurs the least number of migrations in these two figures; Anycast-SP requires a few more migrations than Anycast-MP; Anycast-JRE incurs the most number of migrations; Anycast-EP is comparable to Anycast-JRE in terms of the required number of migrations.
This is why Anycast-EP's performance is close to that of Anycast-JRE. Since migration granularity $\kappa$ is set as $100$ Gbps, the maximum bandwidth capacity of each migration is $100$ Gbps. All these results show that more migrations incur lower cost of the brown energy consumption because migration reduces the brown energy consumption of dirty DCs, and improves the renewable energy consumption utilization efficiency of green DCs.

\section{Conclusion}\label{conclusion}
Inter-DC VM migration brings additional traffic to the network, and the VM mitigation is constrained by the network capacity, rendering inter-DC VM migration a great challenge. This is the first work that addresses the emerging renewable energy-aware inter-DC VM migration problem in the inter-DC network over the elastic optical infrastructure. The RE-AIM problem is a many-manycast communications problem, and the main contribution of this paper is to minimize the total cost of the brown energy consumption of the DCs via VM migration with consideration of the available network resources in an inter-DC network over the elastic optical infrastructure. The RE-AIM problem is formulated as an ILP problem, and proven to be NP-hard. The results are compared with the optimal results for small network configurations. We propose a few heuristic algorithms to solve the large network configurations, and their viabilities in minimizing the cost of the brown energy consumption in inter-DC migration have been demonstrated via extensive simulations.



\appendix

\section{Proof of the RE-AIM problem is NP hard}
The RE-AIM problem includes migrating, routing, and spectrum assignment (MRSA). Here, migrating refers to finding a feasible destination DC for migration, and RSA is required to find a path and assign a (or a few) spectrum slot(s) for each request. We reduce any instance of multiple knapsack problem into the RE-AIM problem.

A multiple knapsack problem can be defined as follows: Given a set of items $X=\{1, 2, ... x\}$ and a set of knapsacks $Y=\{1, 2, ..., y\}$; each item $x_{i}$ is characterized by a weight $xw_{i}$, a volume $xo_{i}$ and a value $xa_{i}$; each knapsack is limited by a volume capacity $yv_{i}$ and a weight capacity $yw_{i}$ $(x_{i}>0, y_{i}>0, xw_{i}>0, xo_{i}>0, xa_{i}>0, yv_{i}>0, yw_{i}>0,\quad\forall i)$ \cite{multiple_knapsack_2000}. The objective is to place as many items as possible in all knapsacks, without exceeding the limit of each knapsack, such that the total value of items placed in all knapsacks is maximized.

For the RE-AIM problem, a request $r$ is mapped into an item $x_{i}$ and each DC into a knapsack $y_{i}$. The computing resource and the bandwidth resource requirements for the request $r$ are mapped into the volume $xo_{i}$ and the weight $xw_{i}$, respectively. The cost of the brown energy of the $r$th request is mapped into the value $xa_{i}$. The maximum available computing resource in DC $m$ is mapped into the volume capacity $yv_{i}$, and the maximum available spectrum slots in the EON into the weight capacity $yw_{i}$. The objective is to minimize the cost of brown energy, which is the same as the objective of maximizing the green energy utilization according to the cost of brown energy. Without considering the routing and spectrum assignment, any instance of multiple knapsack can be reduced into the RE-AIM problem. Since the multiple knapsack problem is NP-hard \cite{multiple_knapsack_2000}, the RE-AIM problem is also NP-hard. Additionally, the routing and spectrum assignment problem is also proved to be an NP-hard problem \cite{Christodoulopoulos2011}. As a result, the RE-AIM problem is NP-hard.



%

\bibliographystyle{IEEEtran}

\end{document}